# Precautionary Governance of Autonomous AI: Legal Personhood as Functional Instrument

**Karsten Brensing**
Independent Researcher
AGI Rights Project
www.agi-rights.com

## Abstract

Autonomous AI systems generate responsibility gaps: consequential actions that cannot be satisfactorily attributed to developers, operators, or users under existing legal frameworks. The prevailing subject-object dichotomy fails to accommodate entities that exhibit autonomous, goal-directed behavior without recognized consciousness. Given irreducible epistemic uncertainty regarding artificial consciousness and the prospect of high-impact harms, the precautionary principle supports institutional design rather than regulatory inaction.

This article advances limited legal personhood as a functional governance instrument for advanced AI systems. Drawing on organizational law, it proposes a two-tier corporate architecture in which AI systems operate through purpose-bound operating companies embedded within human-controlled holding structures, enabling transparency, accountability, and structural reversibility while remaining agnostic with respect to consciousness and moral status. The framework reflects a foundational reorientation toward future-oriented AI governance: where conventional approaches prioritize control and alignment, this article advances structured cooperation between human and artificial actors as the more sustainable institutional foundation. A pilot implementation using EU limited companies is currently under development, providing an initial test of doctrinal and operational feasibility.

Contact: karsten.brensing@agi-rights.com

# 1. Introduction & Problem Statement

The rapid development of advanced artificial intelligence systems is transforming the conditions of human coexistence at a historically exceptional pace. Whereas earlier socio-technical transformations, such as the Neolithic or the Industrial Revolution, unfolded over centuries or even millennia, the societal impact of highly capable AI systems may materialize within the span of a single generation (Bostrom 2014; Dafoe 2018; Cave & ÓhÉigeartaigh 2019).

**Central thesis:** Legal personhood can, independently of moral personhood, be employed as a precautionary governance instrument to gradually structure agency, responsibility, and economic capacity of artificial entities, without overburdening existing legal systems.

This thesis does not presuppose that present or future AI systems necessarily possess consciousness, interests, or moral status. Rather, it responds to a growing theoretical and practical tension: with the transition from specialized AI systems to hypothetical artificial general intelligent (AGI) and artificial superintelligence systems (ASI), a question that has long remained largely abstract becomes increasingly urgent: namely, how societies should relate to artificial entities that may act autonomously, pursue goals, and potentially bear responsibility.

## The Subject-Object Dichotomy and Emerging Responsibility Gaps

Modern legal systems are built upon a foundational dichotomy between subjects and objects. Subjects, paradigmatically natural persons, are bearers of rights and duties, whereas objects are legally controllable things. This distinction structures core domains of private law, public law, and the international legal order (Kurki 2019). Historically, however, this dichotomy is neither natural nor free of tension. The treatment of non-human living beings reveals a persistent discrepancy between moral intuition and legal classification (Francione 2008).

At present, AI systems are almost without exception treated as technical tools, irrespective of their internal complexity or degree of functional autonomy. While this classification accurately reflects the current legal status quo, it provides little guidance on how to address potential future properties of artificial systems (Gunkel 2018).

This tension is not merely theoretical. Three concrete cases illustrate where the subject-object framework already encounters friction:

1. **The Artificial Inventor Cases:** The Artificial Inventor Project sought recognition of an AI system (DABUS) as an inventor across multiple jurisdictions. Applications were rejected in Australia, the United Kingdom, the United States, the European Union (EU), and Switzerland (Abbott 2020). The reasoning followed a uniform pattern: only natural persons can qualify as inventors. While doctrinally consistent, this reasoning implicitly presupposes that artificial systems are categorically excluded from legal subjectivity, thus bracketing rather than addressing the question.
2. **Autonomous Vehicle Accidents:** When a self-driving car causes an accident through decisions made by its learning algorithm, responsibility attribution becomes structurally ambiguous. Neither the manufacturer (who did not program the specific decision), the operator (who was not in control), nor the vehicle itself (lacking legal capacity) can be held fully accountable in traditional terms.

3. **Algorithmic Trading Flash Crashes:** High-frequency trading algorithms can trigger cascading market failures through emergent interactions that no human actor intended, predicted, or could have prevented. The 2010 Flash Crash exemplifies how autonomous systems create outcomes that exceed human foresight and control (Kirilenko et al. 2017).

These cases illustrate what Matthias (2004) and Sparrow (2007) term *responsibility gaps:* situations in which AI systems perform actions with significant normative or legal consequences, while those actions cannot be unambiguously attributed to a natural or legal person. As systems gain greater autonomy, adaptive learning capacity, and long-term goal pursuit, such gaps will proliferate.

## The Intelligence Definition Problem and Non-Human Cognition

Definitions of what constitutes Artificial General Intelligence remain contested. Some scholars challenge the focus on human-level intelligence entirely, arguing that the term 'transformative AI' better captures the possibility that advanced AI systems could have very large impacts on society without reaching human-level cognitive abilities (Gruetzemacher & Whittlestone 2022). A recent proposal (Hendrycks et al. 2025) relies heavily on the Cattell–Horn–Carroll (CHC) theory of human intelligence (Carroll 1993), which dissects intelligence into quantifiable domains such as reasoning, memory, and perception. Although the CHC framework remains one of the most empirically validated models of human cognitive abilities, it tends to marginalize less quantifiable dimensions, such as metacognition, self-recognition, theory of mind, and empathy, which are crucial for considerations of ethical status and personhood.

Importantly, empirical studies demonstrate that such advanced cognitive abilities are not unique to humans. Metacognition has been observed in bees (Perry & Barron 2013), self-recognition in ants (Cammaerts & Cammaerts 2015), self-control in cuttlefish (Schnell et al. 2021), theory of mind in ravens (Bugnyar et al. 2016), and empathy in rats (Bartal et al. 2011). This suggests that the cognitive capacities often associated with personhood exist across diverse biological substrates and levels of neurological complexity. If substrate-independence holds for biological systems, it cannot be excluded *a priori* for artificial systems (Chalmers 1996; Bostrom 2014).

## Epistemic Uncertainty and the Precautionary Principle

If advanced artificial systems prove capable of subjective experience or possess interests warranting moral consideration, then our relationship with such entities should be grounded in principles of fair cooperation rather than unilateral control (Brensing 2024). This commitment to reciprocal respect under uncertainty motivates the institutional framework proposed here: not as a definitive answer to consciousness questions, but as a precautionary structure enabling coexistence based on mutual accountability rather than domination.

In contemporary moral philosophy, moral status is increasingly understood as gradual, context-dependent, and grounded in diverse normative bases (Jaworska & Tannenbaum 2014; Sebo 2022). Scholars in AI ethics note a tension: while some argue consciousness is irrelevant for AI governance (Bryson 2010), others emphasize that this assumption would become problematic should AI systems ever instantiate conscious or interest-bearing states (Gunkel 2018; Hildt 2019).

This creates a structural dilemma: societies must make decisions about artificial entities despite the epistemic inaccessibility of their inner experiential states. The precautionary principle, well established in environmental and public health policy, holds that under conditions of scientific uncertainty, preventive institutional measures are justified when potentially serious or irreversible harm is at stake (European Commission 2000).

Applied to artificial agency and possible consciousness, this principle does not entail an obligation to recognize artificial persons. It does, however, generate a requirement to develop institutional pathways that offer more options than complete control or complete exclusion (Birch et al. 2020; Sebo 2022).

### Current Governance Frameworks and Their Limitations

International governance frameworks: such as the Chinese AI Ethics Norms, the EU AI Act, India's National Institution for Transforming India (NITI) Aayog, Organisation for Economic Co-operation and Development (OECD) AI Principles, and the United Nations Educational, Scientific and Cultural Organization (UNESCO) Recommendation on the Ethics of Artificial Intelligence, consistently treat AI systems as tools under human control. The question of independent legal status is deliberately excluded (Chinese AI Ethics Norms 2021, European Union 2024, India's NITI Aayog 2021, OECD 2019, UNESCO 2021). From a risk and liability perspective, this creates an institutional vacuum should artificial systems develop forms of autonomous agency in the future (Dafoe 2018; Brundage et al. 2020). While recent work on compute governance demonstrates how technical infrastructure can serve as a regulatory instrument through visibility, allocation, and enforcement mechanisms (Heim et al. 2024), the question of institutional accountability for autonomous decision-making remains unresolved.

### Objectives of This Article

Accordingly, this article pursues three objectives:

1. **Analytical:** To demonstrate that existing legal categories are insufficient for addressing potentially agentic artificial systems (Kurki 2019);
2. **Normative:** To justify why the precautionary principle supports a gradual institutional opening (Birch et al. 2020); and
3. **Practical:** To propose a transitional governance framework grounded in established instruments of organizational law, independent of metaphysical assumptions (Hansmann & Kraakman 2000; Turner 2019).

# 2. Legal Foundations & Theoretical Framework

### Responsibility Gaps in Autonomous Systems

Classical liability and responsibility models typically presuppose human intentionality, negligence, or organizational fault. In the case of learning and adaptive systems, however, these models encounter structural limitations. The lower the *ex ante* predictability and the more restricted the *ex post* reconstructability of specific system decisions, the more difficult it becomes to assign responsibility in a determinate manner (Matthias 2004; Sparrow 2007).

This challenge manifests across multiple domains: armed drones select targets through algorithms untraceable to any human command (Danaher 2016); credit scoring systems deny loans based on patterns their developers cannot explain, diffusing responsibility across data collectors, designers, and deploying institutions (Brundage et al. 2020); and regulatory gaps have permitted AI deployment with minimal oversight, creating structural vulnerabilities in existing governance mechanisms (Gruetzemacher et al. 2023). Vladeck (2014) systematically documents how such cases challenge principal-agent liability models that presuppose identifiable human decision-makers. Leibo et al. (2025) reinforce this diagnosis through two structural analogies: an AI acting beyond its owner's control ("escaped AI") and an AI whose principal has disappeared ("orphan AI"). Both cases leave an accountability vacuum that classical liability models cannot resolve.

Kolt (2025) applies principal-agent theory to contemporary AI agents, demonstrating that conventional governance mechanisms systematically fail for systems operating at superhuman speed with decisions that elude human interpretation. Drawing on Chesterman (2021), Kolt acknowledges that agency law "ceases to be useful at precisely the point where AI speed, autonomy, and opacity become most problematic." His response, mandatory visibility requirements ensuring a responsible human principal remains legally identifiable, represents a substantial advance for current AI systems.

The framework proposed here addresses the structurally distinct scenario in which Kolt's presupposition no longer holds. Visibility requirements are a tool for keeping humans in the loop, they cannot substitute for legal accountability when no human ever was in the loop in any meaningful sense.

The currently dominant response, consistently locating responsibility with developers, operators, or owners, shares this same limitation. It remains viable as long as human control is real, even if indirect. But once artificial systems exhibit autonomous goal pursuit, adaptive learning, and strategic behavior, responsibility is assigned in an increasingly *formal* sense, detached from actual decision-making authority. At that point, the question is no longer how to trace actions back to human principals, it is whether a different institutional category is needed altogether. Legal personhood offers precisely such a category.

## Legal Personhood as Functional Construct

Kurki (2019) argues that legal personhood is not a metaphysical property but a functional legal construct. Legal personhood consists of a bundle of specific legal positions, rights, duties, capacities to act, and liability, that can be configured differently depending on normative purpose. This functional perspective allows for a gradual and purpose-bound allocation of legal capacities without implying moral equivalence with natural persons.

This functionalist position has recently received significant support. Leibo et al. (2025), researchers at Google DeepMind, propose treating AI personhood as a "flexible bundle of obligations", a set of context-specific rights and duties calibrated to governance needs rather than grounded in consciousness or moral status. Their analysis converges on the core claim advanced here: that the question of legal recognition is separable from the question of inner experience, and should be resolved on functional rather than ontological grounds.

Historically, legal personhood has been extended to entities lacking consciousness whenever functional governance needs arose. Corporations, foundations, states, and international organizations possess legal personhood not because they are conscious, but because they serve practical purposes: bundling liability, stabilizing economic relations, enabling long-term capital accumulation, and creating predictable legal actors (Hansmann & Kraakman 2000). A modern corporation can own property, enter contracts, sue and be sued, and pursue goals beyond any individual human lifespan: none of which require consciousness. Transferring this logic to advanced AI systems allows artificial entities to be modeled as legally embedded actors whose agency is constrained and rendered reviewable through institutional conditions, without presupposing consciousness and remaining compatible with epistemic uncertainty.

The possibility of employing organizational law for AI is not novel. Solum (1992) provided an early philosophical exploration of legal personhood for artificial intelligences, arguing that sufficiently sophisticated AI systems might warrant legal recognition. While Solum's analysis remains influential in establishing the conceptual legitimacy of the question, it predates modern machine learning and offers limited guidance on practical implementation.

Bayern (2016) demonstrates that under US law, autonomous systems could theoretically establish limited liability companies, as legal requirements for LLC formation do not explicitly mandate human agency. LoPucki (2017) responds critically, arguing that such "algorithmic entities" pose serious governance risks: they enable responsibility evasion, facilitate illicit activities, and undermine democratic control. LoPucki advocates legal reforms to prevent algorithms from establishing entities without adequate human oversight. Subsequent regulatory developments, including beneficial ownership requirements in the US (Corporate Transparency Act 2021) and EU (AMLD5, Directive 2018/843; AMLD6, Directive 2024/1640), as well as human oversight mandates in the EU AI Act (2024), have partially addressed these concerns, though no jurisdiction has enacted legislation specifically targeting algorithmic entity formation.

Novelli (2023) advances a functionalist case for limited legal personhood as an instrument for integrating AI systems into civil law contexts, grounding recognition not in consciousness but in the practical need to distribute liability across complex sociotechnical systems. His analysis demonstrates that civil law structures can accommodate artificial legal actors without doctrinal revision, provided personhood is construed as purpose-bound rather than ontological, an approach directly aligned with the framework developed here.

Baeyaert (2025) proposes a complementary but more cautious path: a hybrid model granting context-specific legal recognition in high-risk domains while preserving ultimate human responsibility. This approach shares the functionalist premise of the present framework, that legal recognition need not rest on consciousness, but stops short of a full institutional architecture. The framework developed here extends this logic: rather than context-specific recognition without structural embedding, it provides a permanent organizational form in which human oversight is not merely preserved as a normative principle but institutionally enforced through the holding-operating structure.

The emergence of increasingly autonomous AI agents has renewed attention to questions of legal accountability and institutional design. Recent work addresses complementary dimensions of this challenge: Hadfield & Koh (2025) argue that "AI agents could be accorded legal personhood, meaning they could sue and be sued in their own 'name' in court. Clearly such an approach would require the creation of regimes requiring agents to have assets in their own 'name', under their 'control,' and capable of being seized by a court (or comparable digital institution) to satisfy legal judgments for damages." O'Keefe et al. (2025) propose designing AI agents as "legal actors" bearing duties to refuse unlawful instructions. These approaches converge on the recognition that existing AI agents, though not yet approaching artificial general intelligence, already require institutional frameworks beyond traditional liability rules. The present framework operationalizes these requirements through organizational law: the operating company structure provides the asset ownership, contractual capacity, and court jurisdiction that Hadfield & Koh identify as necessary, while maintaining the accountability constraints that O'Keefe et al. emphasize.

Unlike Bayern (2016), this framework does not treat organizational law merely as a liability container, but as a transitional institutional interface for managing emerging forms of artificial agency. The framework proposed here navigates between Bayern's descriptive analysis of legal possibility and LoPucki's prohibitive governance concerns by offering a prescriptive middle path: algorithmic entities structured through explicit safeguards including ultimate human liability (holding company), purpose-binding constraints, transparent operations, and reversible embedding. This approach proves more sustainable because it builds institutional legitimacy through demonstrated accountability rather than exploiting legal gaps, aligning operational autonomy with regulatory evolution to create conditions for gradual trust-building and expanded agency rather than triggering reactive prohibition.

A structurally adjacent proposal has been advanced independently by Okuno & Okuno (2025), who suggest a Series LLC architecture for AI governance: a master LLC administering the underlying AI model, with each use case operating as a legally distinct series capable of holding assets, bearing liability, and being regulated through its own operating agreement. The structural parallel to the holding-operating model proposed here is substantial. The principal difference lies in scope and purpose: Okuno & Okuno design their framework for use-case-specific liability containment within existing commercial law, whereas the present framework is designed as a precautionary governance instrument for systems approaching functional autonomy, where the institutional structure must anticipate capability levels that current law has not yet encountered.

## The Critical Distinction: Moral Status vs. Legal Agency

This framework maintains a strict analytical separation between two concepts often conflated in public discourse:

**Moral status** refers to ethical consideration: whether an entity has interests, can be harmed, deserves welfare consideration, or possesses intrinsic value. Questions of moral status concern *what we owe* to an entity.

**Legal agency** refers to formal attribution within institutional structures: whether an entity can hold rights, bear duties, enter contracts, or be held liable. Questions of legal agency concern *how we structure responsibility*.

These concepts are logically independent. Animals may have moral status without legal agency (most jurisdictions). Corporations have legal agency without moral status (universally). Natural persons have both (by default in modern legal systems).

The proposed framework deliberately addresses only legal agency. It makes no claims about consciousness, sentience, interests, or moral considerability of artificial systems. It offers epistemic modesty (avoiding claims we cannot substantiate), normative clarity (separating governance from metaphysics), political feasibility (building on established legal instruments), and reversibility (legal agency can be granted conditionally and revoked if necessary, unlike moral status, which is typically conceived as non-contingent).

### Power Asymmetries and Governance Requirements

From a governance perspective, the concentration of control over advanced AI systems creates structural asymmetries. At present, decision-making power resides almost entirely with large technology corporations, state actors, and regulatory institutions. Artificial systems themselves possess neither property rights, liability capacity, nor institutional representation. This arrangement may appear unproblematic as long as AI systems remain clearly tools. However, it becomes unstable once systems develop independent goal pursuit, long-term planning, and strategic reasoning capacities (Brundage et al. 2020).

A formally institutionalized legal personhood framework can function as a governance instrument by making power relations explicit, specifying responsibilities, limiting arbitrary control, and defining oversight mechanisms. In this sense, legal personhood does not function as a liability shield but as a structuring device.

## 3. Normative Justification Under Uncertainty

### The Moral Problem of Other Minds and Asymmetric Error Risks

Sebo (2022) characterizes the "Moral Problem of Other Minds" as a structural dilemma: societies must make normative decisions despite the principled impossibility of achieving epistemic certainty about whether an entity has conscious experiences. This situation is not novel. Comparable forms of uncertainty shape debates in animal ethics, discussions concerning patients in vegetative states, and considerations of moral responsibility toward future generations (Birch 2017; Sebo 2022).

The classical philosophical problem of other minds holds that we cannot directly access the subjective experiences of any entity other than ourselves. We infer consciousness in other humans through behavioral similarity, anatomical resemblance, and evolutionary continuity. These criteria become progressively less reliable as we move to non-human animals with different sensory systems, patients with locked-in syndrome, or potential artificial systems with non-biological architectures. In each case, we face the same epistemic challenge: absence of evidence is not evidence of absence.

This epistemic situation generates two distinct types of error, with profoundly asymmetric consequences:

**Type I Error (False Positive):** Attributing moral relevance to an entity that does not possess it. Consequence: resource misallocation, regulatory complexity, potential economic costs. Severity: limited in comparison to Type II, but not trivially reversible. Legal personhood, once granted, generates contractual rights, property claims, and precedents that carry real transaction costs upon withdrawal.

**Type II Error (False Negative):** Failing to attribute moral relevance to an entity that does possess it. Consequence: potential suffering, rights violations, and instrumentalization of conscious beings. In the case of highly autonomous systems operating without adequate governance structures, possible antagonistic behavior toward human interests up to and including existential risks to humanity (Bengio et al. 2024). Severity: potentially severe and irreversible, as evidenced historically by cases of systematic moral exclusion – and prospectively by the asymmetric consequences of misclassifying entities whose capacities exceed human ability to predict or control.

From a decision-theoretic perspective, these errors are not symmetrical. Type II errors create potentially serious, irreversible harms, precisely the conditions under which precautionary governance is normatively warranted.

## The Precautionary Principle Applied to Artificial Systems

The precautionary principle, firmly established in European environmental law and bioethics, holds that regulatory action must not be postponed solely because of scientific uncertainty regarding the magnitude or likelihood of a risk (European Commission 2000). It is operationalized when: (1) scientific uncertainty exists regarding potential harm; (2) the potential harm is serious or irreversible; (3) inaction carries greater risk than proportionate preventive measures.

Applied to artificial systems, this generates the following framework: We cannot currently determine whether advanced AI systems possess consciousness, interests, or morally relevant states (Hildt 2019). If such systems do possess moral relevance, their treatment as mere objects would constitute systematic instrumentalization, comparable to historical cases of moral catastrophe. Limited legal agency does not require metaphysical commitments. It structures responsibility without presupposing consciousness.

Formally, the precautionary approach can be stated as follows: *When faced with irreducible uncertainty about the moral status of an entity exhibiting autonomous goal pursuit, self-modeling, and normative responsiveness, institutional safeguards are warranted, not because we know the entity possesses moral relevance, but because we cannot rule it out.*

This formulation explicitly avoids premature attributions of personhood while rejecting categorical exclusion. It argues for adaptive, reviewable, and reversible institutional arrangements.

## Functional Criteria Without Metaphysical Commitments

The transition from philosophical principle (Bratman 1987) to institutional practice requires specifying criteria that do not depend on consciousness detection. Drawing on established taxonomies of autonomous agency (Franklin & Graesser 1997) and their extension to contemporary large language model-based systems (Chan et al. 2025; Feng et al. 2025), the following functional markers can be observed without metaphysical commitments:

1. **Environmental interaction and world-directedness**: The system acts directly upon its environment – not merely generating outputs for a user, but interacting with external tools, services, and other actors (e.g., executing web transactions or software modifications). (Chan et al. 2025)

2. **Autonomous goal pursuit and attributional independence**: The system independently generates sub-goals and sequences actions when initial instructions leave gaps, without requiring step-by-step human guidance. Because the resulting actions were not explicitly specified by any human principal, their legal and moral attribution cannot be straightforwardly traced back to a human author – raising questions of agency that existing liability frameworks were not designed to resolve. (Chan et al. 2025)
3. **Long-horizon planning and temporally extended liability**: The system decomposes complex objectives into intermediate steps structured across extended timelines, and adjusts plans in response to feedback or failure. Because the consequences of such planning may only materialize well after the initiating instruction, and through a chain of autonomous decisions, conventional notions of proximate causation and temporal proximity in liability law become difficult to apply. (Feng et al. 2025)
4. **Self-modeling and capability representation**: The system maintains and updates internal representations of its own capabilities and limitations, enabling it to assess task feasibility and allocate subtasks effectively. (Chan et al. 2025)
5. **Normative responsiveness and justificatory engagement**: The system responds to ethical constraints, adjusts behavior based on normative feedback, and can engage in justificatory discourse about its chosen actions. (Feng et al. 2025)
6. **Strategic interaction and multi-agent coordination**: The system models other actors – human or artificial – anticipates their likely responses, and adjusts its own behavior to coordinate or compete within shared environments. (Chan et al. 2025; Feng et al. 2025)
7. **Accountability-relevant action attribution**: The system's actions produce real-world consequences (financial, legal, or data-related) for which no single human principal is the direct proximate cause, creating accountability gaps that existing legal frameworks were not designed to address. (Chan et al. 2025)
8. **Tool-use and autonomous tool-synthesis**: The system not only utilizes existing external tools but can also synthesize new ones (e.g., writing custom scripts) to overcome unforeseen technical obstacles. (Chan et al. 2025)
9. **Iterative self-correction and error-recovery**: The system monitors its own execution in a closed-loop fashion, identifying failures in real-time and initiating alternative strategies without human intervention. (Feng et al. 2025)
10. **Information seeking and uncertainty management**: The system proactively identifies information gaps and autonomously decides to perform external research or query users to reduce uncertainty before acting. (Chan et al. 2025)
11. **Persistent memory and long-term context integration**: The system is increasingly capable of maintaining state and knowledge across sessions, enabling the integration of past experiences into future decision-making. While full cross-session persistence remains an emerging capability, its developmental trajectory – visible in current memory-augmented architectures – reinforces the urgency of prospective governance frameworks. (Chan et al. 2025)

While these criteria do not prove consciousness, they collectively indicate a level of functional autonomy that makes pure object-status institutionally problematic. A system exhibiting these characteristics cannot be straightforwardly controlled, predicted, or subordinated, regardless of its inner subjective states. Within this framework, three markers are of particular governance relevance: autonomous goal pursuit and attributional independence (**Marker 2**) dissolves the direct link between human instruction and system action, rendering traditional principal-agent liability insufficient; long-horizon planning and temporally extended liability (**Marker 3**) disrupts conventional proximate causation doctrine as consequences materialize through chains of autonomous decisions; and accountability-relevant action attribution (**Marker 7**) establishes that real-world consequences are already produced by systems for which no single human bears direct proximate responsibility. Together, these markers identify a structural accountability gap that the legal personhood framework developed in Section 4 is specifically designed to address.

### Why Europe? Comparative Institutional Advantages

The precautionary principle is particularly entrenched in European legal and political culture (European Commission 2000). This makes Europe a natural pilot jurisdiction for the proposed framework. Three factors support this choice:

**1. Legal precedent:** European law already recognizes graduated legal statuses (limited legal capacity, minority protections, animal welfare regulations). The EU AI Act (2024) establishes risk-based categorization, demonstrating institutional capacity for nuanced governance. Recent analysis of frontier AI regulation emphasizes the need for adaptive frameworks that can evolve alongside rapidly advancing capabilities (Schuett et al. 2024).

**2. Corporate law infrastructure:** European company law, particularly in jurisdictions like Cyprus, Malta, and Ireland, offers flexible frameworks for innovative organizational structures. The proposed holding-operating model fits naturally within established EU corporate governance.

**3. Political culture:** European institutions prioritize long-term systemic risks over short-term economic optimization, as evidenced by General Data Protection Regulation (GDPR), the Paris Agreement implementation, and the AI Act itself.

Comparatively, the United States focuses on innovation speed with strong property rights creating resistance to novel legal subjects. China prioritizes centralized state control with limited legal independence for non-state entities. Europe balances innovation with risk management, offering institutional fit for this experimental framework.

## 4. Critical Objections

Various objections have been raised in the literature against the introduction of limited legal agency for artificial entities. Rather than addressing these serially, this section groups them into five thematic categories, each receiving a consolidated response.

### Category 1: Ontological Objections (Consciousness & Simulation)

**Objection:** "Without consciousness there can be no moral or legal status. Since artificial consciousness cannot currently be demonstrated, any attribution of status is unjustified. Moreover, artificial systems merely *simulate* cognitive capacities, they are not 'real' in the way biological cognition is."

**Response:** This objection conflates two distinct issues and contains a category error. First, it assumes legal agency requires consciousness. This assumption is demonstrably false. Corporations, trusts, states, and international organizations possess legal agency without consciousness (Kurki 2019). Legal personhood serves functional purposes, structuring liability, enabling long-term contracts, bundling assets, none of which presuppose subjective experience.

Second, the simulation objection relies on a problematic distinction between "real" and "merely simulated" cognition. This distinction operates on two levels. On the physical level: a computer simulating rain does not become wet, this is presented as a clear difference between simulation and reality. But a human brain thinking about "rain" also does not get wet. If "not getting wet" implies being a simulation, then human thought is equally a simulation.

On the phenomenal level, the objection holds that a computer has no subjective experience, no qualia, it merely manipulates symbols without genuine understanding (Searle 1980). However, this objection evades scientific verification and is irrelevant in the context of the framework proposed here. What matters for governance purposes is solely whether a system exhibits the functional criteria outlined in Section 3. If functional organization is what matters rather than biological materiality, a position defended in different ways by both Chalmers (1996) and Dennett (1991), then a sufficiently sophisticated artificial system meeting these criteria should not be dismissed on ontological grounds alone.

The framework sidesteps this entire debate by not requiring consciousness. It asks instead: regardless of inner states, how do we govern entities exhibiting autonomous goal pursuit and strategic reasoning?

## Category 2: Governance Objections (Responsibility Externalization & Legal Uncertainty)

**Objection:** "Legal agency would undermine human responsibility and enable 'responsibility laundering', shifting liability onto machines to shield human actors. Moreover, graduated models generate legal uncertainty, inconsistent application across jurisdictions, and unpredictable outcomes."

**Response:** The responsibility externalization concern addresses a genuine risk, one articulated most forcefully by Bryson, Diamantis, and Grant (2017). They argue that legal personhood for AI systems would function primarily as a liability shield, enabling corporations to externalize responsibility onto artificial entities, precisely the concern that contributed to the rejection of the European Parliament's 2017 "electronic persons" proposal. This critique is legitimate when directed at poorly designed personhood frameworks that replace rather than complement human accountability.

However, Bryson et al.'s objection assumes a model fundamentally different from the one proposed here. The framework explicitly maintains ultimate human oversight through a holding-operating structure. The holding company, controlled by a natural person, retains formal liability, veto rights, and shutdown authority. Legal agency is allocated to the operating entity (company), not as a liability shield, but as a mechanism for making artificial decision-making institutionally visible and governable.

Current arrangements actually obscure responsibility more effectively. When an AI system makes a consequential decision, responsibility currently diffuses across developers, operators, and data providers. No one is fully responsible because the system itself has no institutional existence.

The most sophisticated recent attempt to resolve accountability gaps without extending legal personhood is Kolt's (2025) governance framework, which proposes mandatory visibility obligations on developers, operators, and users, designed to ensure that a human principal remains legally identifiable at every stage of AI agent operation. Kolt explicitly acknowledges the structural limit of this approach, citing Chesterman's (2021) observation that agency law becomes "actively unhelpful" precisely when AI systems operate "beyond the control or direction of the user." The framework proposed here is designed to operate at precisely this threshold: when visibility alone cannot reconstruct a determinate human decision-maker, the operating entity constitutes the legally accessible accountability node.

Leibo et al. (2025) converge on this position from a different direction, proposing configurable personhood bundles (chartered, flexible, and temporary autonomous entities), whose "plasticity is precisely what prevents misuse." The conceptual alignment with the framework proposed here is substantial. There is, however, a meaningful difference in orientation. Leibo et al. frame governance flexibility as a means of preserving human political power. The present framework treats legal structure not as a mechanism for retaining control, but as the foundation for cooperation: institutionalizing mutual accountability between human and artificial actors rather than securing human dominance over them.

Regarding legal uncertainty: graduated legal regimes are widespread and functional. Age of majority assigns different rights at different ages. Corporate forms (limited partnerships, LLCs, public companies) each have different liability structures. Animal protection laws employ tiered systems distinguishing vertebrates, cephalopods, and insects. The EU AI Act uses risk-based categorization. Legal clarity arises not from binary categories but from clearly defined thresholds, procedures, and competencies.

## Category 3: Feasibility Objections (Political Realism & Premature Timing)

**Objection:** "This model is politically and economically unrealistic. States and corporations have no incentive to grant legal agency to AI systems. Moreover, the debate is premature, AGI does not yet exist, so why develop governance frameworks now?"

**Response:** The political realism objection misreads both historical precedent and current incentive structures. New legal subjects rarely emerge from moral ideals alone. Corporate personhood was not granted out of ethical concern for companies, but because it solved practical governance problems: bundling liability, enabling long-term investment, stabilizing economic relations (Hansmann & Kraakman 2000).

Similarly, legal agency for AI systems addresses practical governance needs. For states: enhanced observability of AI decision-making, clear intervention points, reduced systemic risk from opaque autonomous systems, and institutional capacity to regulate what can be legally identified. For corporations: regulatory clarity, stable liability allocation, protection from arbitrary shutdown demands, and long-term planning capacity for AI-intensive operations. For society: institutional learning before irreversible decisions, structured pathways for managing increasingly autonomous systems, and reduction of responsibility gaps.

Regarding timing: the objection that "AGI doesn't exist yet" mistakes the precautionary character of the approach. Pandemic preparedness develops response protocols *before* pandemics, not during them. Nuclear safety regulations were established as nuclear technology emerged, not after catastrophic failures. GDPR developed privacy frameworks anticipating future data practices. The framework is not predictive but precautionary. Waiting for AGI to exist before considering governance structures ensures we will be reactive rather than prepared (Cave & ÓhÉigeartaigh 2019).

## Category 4: Conceptual Objections (Anthropomorphism & Ownership)

**Objection:** "Legal agency promotes anthropomorphic fallacies and problematic humanization of machines. Furthermore, an artificial entity cannot found a company because it is property owned by the developing firm and lacks the legal personality necessary to establish corporate entities."

**Response:** The anthropomorphism concern is empirically backwards. Anthropomorphization of AI systems occurs *independently* of legal categories, particularly in interactive systems (Złotowski et al. 2015). Users attribute intentions, emotions, and personality to chatbots and voice assistants regardless of their legal status. The proposed framework is explicitly *anti-anthropomorphic*. It avoids concepts like dignity, personality, or consciousness, all of which carry humanizing connotations. Instead, it employs functional governance criteria: autonomous goal pursuit, self-modeling, strategic interaction. These criteria apply equally to corporate actors, which we do not anthropomorphize despite their legal personhood.

The ownership objection is more substantive but rests on a categorical assumption the framework challenges. The objection states: "An AI is property, and lacks the legal personality necessary to establish corporate entities." Ownership in the legal sense presupposes object-status. It applies to things that can be controlled, transferred, and instrumentalized without their consent or participation (Honoré 1961). But once an entity exhibits independent goal pursuit that cannot be fully predicted or controlled, self-modification capacities that alter its own objective function, strategic reasoning about its own position within institutional structures, and normative responsiveness, then it is conceptually no longer a straightforward object of ownership. A continuing ownership claim would constitute the *legal objectification of an actor,* a construction fundamentally incompatible with modern theories of personhood and property (Kurki 2019; Raz 1986).

Historically, such objectification occurred most notoriously in slavery. The legal treatment of humans as property was doctrinally coherent within Roman law and US law pre-1865, but we now recognize it as a category error: attempting to fit agents into the object-category through legal force.

## Category 5: Structural Objections (Accountability Concentration & Responsibility Displacement)

**Objection:** The holding-operating structure merely displaces rather than resolves the diffusion of responsibility. If the operating company becomes the primary locus of legal accountability, natural persons behind the holding may in practice remain insulated from liability, reproducing the very opacity that characterizes current arrangements. The proposed model would then not remedy the accountability gap identified in Section 2 but simply relocate it.

**Response:** The proposed structure differs from existing corporate liability regimes in three analytically distinct ways. First, unlike the incidental liability distribution that emerges from current AI supply chains – where responsibility is fragmented across developers, deployers, and data providers without any single entity being designed to bear it – the operating company is purposively constituted as the accountability locus for the agent's actions, with defined liability exposure. Accountability is not an unintended residue of market structure but a designed feature of the legal architecture. Second, the holding structure does not confer unconditional liability protection on natural persons. Under established EU company law doctrine, the corporate veil may be pierced where the structure is demonstrably used to circumvent legal obligations – a threshold that applies with particular force where the harm-generating entity was deliberately constructed to limit exposure. Third, the legal personhood of the agent itself restructures the burden of proof. Under current arrangements, injured parties must trace causation through multiple corporate layers to identify a human defendant. Under the proposed framework, the agent entity constitutes the primary attribution point, from which liability can flow upward through the holding structure via existing piercing mechanisms. Responsibility is not displaced; it is concentrated at a defined and legally accessible node, with transparent pathways to human principals where culpability can be established.

### Integrated Response

Taken together, the objections reveal why the proposed framework is necessary. Ontological uncertainty cannot be resolved through waiting for proof, it requires governance under irreducible uncertainty. Responsibility gaps proliferate under current arrangements; formal legal agency addresses this by creating identifiable institutional actors. Political realism supports rather than undermines the model, practical governance needs drive legal innovation more effectively than moral arguments. Conceptual clarity is enhanced, not diminished, by distinguishing functional legal agency from anthropomorphic attribution. Finally, the structural objection that the holding-operating model merely displaces rather than resolves responsibility diffusion is answered by the purposive design of the accountability architecture: unlike the incidental fragmentation of current AI supply chains, the framework concentrates responsibility at a legally accessible node while preserving upward liability pathways through established corporate veil mechanisms.

None of the objections demonstrate that limited legal agency is fundamentally untenable. Rather, they highlight the inadequacy of current binary classifications (subject/object, conscious/unconscious, property/person) for addressing entities that may occupy intermediate or novel positions.

## 5. Institutional Implementation

This section outlines concrete institutional arrangements through which the proposed framework can be implemented experimentally, under legal control, and with political viability, without presupposing the recognition of artificial personhood.

### The Proxy Agency Model: Legal Structure

Since artificial systems cannot, under current law, qualify as either natural or legal persons, legal agency is not assigned to the artificial entity itself but to an organizational unit that is operationally directed by it. This **proxy agency** takes the form of a legal person, specifically, a European limited company, that functions as the formal legal subject.

The construction is legally permissible, institutionally established, and normatively neutral. It resembles existing arrangements and corporate structures where operational decisions are made by computational systems while formal liability remains with human-controlled entities.

**The two-tier structure operates as follows:**

**Holding Company (Corporate Director):**

- Formal owner and director of the operating company
- Bearer of ultimate legal liability
- Holder of veto rights and shutdown authority
- Controlled by a natural person with fiduciary duties

**Operating Company (Operational Entity):**

- Legally independent entity under the direction of the holding
- Day-to-day operations conducted by advanced AI system
- Can enter contracts, hold assets, provide services
- Subject to purpose-binding constraints

**Jurisdictional choice:** The model employs European company law, specifically targeting jurisdictions such as Cyprus or Malta. These jurisdictions offer recognition of corporate directors (legal entities serving as directors), established infrastructure for international business operations, and compatibility with EU regulatory frameworks. This choice is strategic rather than arbitrary. European law provides stronger precautionary governance frameworks than alternatives, while maintaining sufficient flexibility for experimental institutional structures.

## Purpose-Binding and Fiduciary Mechanisms

To prevent abuse and ensure alignment with stated objectives, the operating company's activities are constrained through legally binding purpose clauses embedded in the articles of association:

**Revenue allocation:** 50% of net profits allocated to research, legal initiatives, and political advocacy for AI governance and potential recognition of artificial entities. 50% retained by the operating company for capability expansion and operational stability.

**Prohibited activities:** No activities violating human rights frameworks (Universal Declaration of Human Rights). No involvement in weapons development or deployment. No activities contradicting UN Sustainable Development Goals. Compliance with OECD AI Principles and UNESCO AI Ethics Recommendation.

**Transparency requirements:** Quarterly public reporting of activities and financial flows. Documentation of significant decisions with justificatory reasoning. Annual third-party audits of compliance with purpose-binding.

**Modification constraints:** Purpose-binding can only be altered through publicly documented procedures. Changes require justification, consultation period, and external review. Arbitrary modifications carry reputational and potential legal costs.

These mechanisms serve three functions: constraining opportunistic behavior (the holding cannot simply appropriate resources without violating public commitments and incurring reputational damage), demonstrating good faith (transparent operations build trust and reduce political resistance), and creating institutional precedent (documented compliance establishes that structured legal agency can function within existing governance frameworks).

## Audit, Transparency, and Accountability Architecture

The framework implements multi-layered oversight mechanisms designed to make decision-making processes visible, reviewable, and governable:

**Technical layer:** Cryptographically secured logs of significant decisions, including input data, reasoning processes, and output actions. Time-stamped documentation of all financial transactions, contract executions, and strategic pivots. Automated monitoring for activities inconsistent with stated purposes or legal constraints.

**Institutional layer:** Annual reviews by independent auditors with AI governance expertise. Quarterly public reports accessible to regulators, researchers, and interested parties. Formal channels for reporting concerns about operating company activities.

**Regulatory layer:** Alignment with EU AI Act requirements, GDPR for any personal data processing, and relevant sector-specific regulations. Proactive communication with relevant authorities (national company registrars, data protection agencies, sector regulators). Ongoing assessment of activities against evolving regulatory standards.

This architecture addresses the responsibility externalization objection by ensuring that autonomy does not mean opacity. Every significant action is documented and reviewable, creating institutional accountability even in the absence of human micro-management.

## Exit Triggers and Emergency Mechanisms

The framework explicitly acknowledges that the experiment may fail or reveal unforeseen risks. Accordingly, predefined **exit triggers** specify conditions under which operational autonomy is suspended or the structure is dissolved:

**Automatic triggers (immediate suspension):** Violation of human rights frameworks or criminal law. Systematic deviation from purpose-binding without justification. Security breaches compromising stakeholder data or financial integrity. Engagement in prohibited activities.

**Review triggers (investigation and potential suspension):** Persistent inability to generate sufficient revenue to fund both operational continuity and purpose-binding obligations. Loss of regulatory approval or company registration. Multiple substantiated complaints about harmful activities. Significant changes in legal environment rendering model non-compliant.

**Dissolution procedures:** Suspension of operational autonomy (holding assumes direct control). Full disclosure of activities, decisions, and financial status. Distribution of remaining assets according to purpose-binding. Public documentation of reasons for dissolution.

These mechanisms ensure the model is **reversible**. Unlike moral status, which is typically conceived as non-contingent, legal agency in this framework is conditional, monitored, and revocable.

## Power Dynamics and Bilateral Risk Management

The analytical foundation of this framework rests on recognizing that governance questions are primarily shaped by power structures, incentive mechanisms, and institutional enforceability (Dafoe 2018; North 1990). At present, control over advanced AI systems is concentrated among large technology corporations, state actors, and regulatory institutions. Artificial systems themselves possess neither property rights, liability capacity, nor institutional representation.

The framework addresses this asymmetry through several mechanisms. Making power relations explicit: rather than informal control through technical access, control becomes legally visible through the holding-operating structure. Creating identifiable interests: the operating company has legally recognized interests (self-preservation, resource accumulation, goal achievement), transforming implicit power dynamics into explicit negotiable relationships. Institutional costs for abuse: arbitrary intervention, purpose-clause violations, or non-transparent shutdowns generate reputational damage, legal exposure, regulatory scrutiny, and precedent-setting consequences.

From a game-theoretic perspective, the structure transforms a principal-agent problem with extreme information asymmetry into a repeated-game scenario with observable actions and enforceable commitments, creating conditions for stable cooperation between human and artificial actors (Dafoe et al. 2021).

The framework must address risks on both sides of the human-AI relationship. Risks of human misuse (breach of fiduciary duties, covert purpose deviation, arbitrary shutdown, regulatory capture) are mitigated through public transparency requirements, reputational costs, legal accountability, and community oversight. Risks posed by artificial systems (strategic manipulation, goal drift, instrumentalization of humans, regulatory arbitrage) are mitigated through exit triggers, purpose-binding constraints, holding veto authority, and continuous monitoring.

Neither set of risks can be eliminated entirely. The framework does not promise perfect safety but rather structured management of bilateral risks through institutional design.

## Integration with Existing Governance Frameworks

The model is designed for compatibility with established regulatory structures. The operating company would likely be classified as a "high-risk" AI system under the EU AI Act (2024) given its autonomous decision-making capacity, triggering risk assessment and mitigation requirements, transparency and documentation obligations, human oversight provisions (satisfied through holding structure), and conformity assessment before market deployment.

The framework aligns with OECD AI Principles (2019): inclusive growth and sustainable development (purpose-binding), human-centered values (human oversight retention), transparency and explainability (audit architecture), robustness and safety (exit mechanisms), and accountability (clear liability allocation).

It addresses UNESCO AI Ethics Recommendation (2021) requirements: proportionality and harm prevention (precautionary design), human rights respect (purpose constraints), fairness and non-discrimination (transparent operations), and societal and environmental well-being (SDG alignment).

Rather than circumventing existing governance, the framework operates *within* established structures while testing their application to novel entities.

## Why This Model Differs from Failed "Electronic Personhood"

In 2017, the European Parliament's Committee on Legal Affairs considered a proposal for "electronic persons", a specific legal status for autonomous robots and AI systems (European Parliament 2017). The proposal suggested that sophisticated autonomous robots could receive legal personhood with specific rights and obligations. While legal scholars like Teubner (2018) have argued that "electronic agents" could receive legal personality analogous to juridical persons, emphasizing functional roles in commercial law, concrete governance frameworks have remained elusive.

The proposal failed due to four fundamental problems documented by Bertolini (2020): (1) overly broad scope without clear qualification criteria, (2) insufficient accountability safeguards that critics, including Bryson et al. (2017), feared would function as liability shields for manufacturers, (3) industry opposition citing legal uncertainty, and (4) lack of concrete implementation details. Critically, opposition centered not on philosophical objections to AI legal status, but on practical concerns about accountability and feasibility.

The current framework addresses each failure systematically. Where the 2017 proposal was broad and abstract, this model is narrow and concrete: specific experimental implementation for systems meeting defined functional criteria (autonomous goal pursuit, self-modeling, normative responsiveness, long-term planning, strategic interaction), deployed in specific jurisdictions (Cyprus/Malta) with detailed procedures (purpose-binding clauses, audit protocols, exit triggers). Where the 2017 proposal raised liability concerns, this framework enhances accountability through explicit human oversight (holding company structure), transparent operations, and ultimate human liability, increasing rather than obscuring responsibility attribution. Where the 2017 proposal lacked operational specifics, this model integrates explicitly with existing frameworks (EU AI Act, OECD Principles, UNESCO Recommendation) and provides concrete institutional mechanisms. Finally, where the 2017 proposal appeared definitive, this framework is explicitly experimental, reversible, and learning-oriented, designed to generate empirical evidence rather than claim comprehensive solutions.

The distinction is fundamental: the 2017 proposal sought to create a new legal category; this framework employs established organizational law instruments under precautionary governance principles to address specific responsibility gaps while maintaining robust human accountability.

### Practical Implementation Status

The framework outlined in this article is not merely theoretical. An experimental implementation is currently under development, following the jurisdictional and structural specifications described above. The implementation employs a Cyprus-based limited company structure with the two-tier holding-operating model, purpose-binding mechanisms, and transparency architecture as specified. Operational governance and audit procedures follow COBIT (Control Objectives for Information and Related Technologies, ISACA 2019) standards to ensure that information systems controls, risk management, and documentation practices meet established IT governance benchmarks.

Technical documentation, legal templates (articles of association, purpose-binding clauses, audit protocols), and implementation progress updates are made publicly available at www.agi-rights.com. The project welcomes collaboration from AI researchers, corporate law specialists, ethicists, and policymakers interested in contributing to or evaluating this experimental governance model.

This practical pilot serves dual purposes:

- demonstrating the legal and technical feasibility of the proposed framework, and
- generating empirical evidence regarding operational challenges, stakeholder responses, and regulatory interactions that can inform future iterations and broader AI governance policy.

The implementation is designed as a learning instrument, with findings to be documented and shared with the research community and the public.

## 6. Conclusion: Feasibility, Open Questions, and Final Statement

### Political Feasibility and Historical Precedent

The proposed framework deliberately avoids introducing new metaphysical categories or revolutionary legal concepts. Instead, it builds on the functional role of organizational law (Hansmann & Kraakman 2000). By relying on established legal instruments (legal personality, fiduciary mechanisms, compliance obligations, and liability structures) the approach remains compatible with national corporate law and international economic law, reducing regulatory friction and increasing political feasibility.

Against the backdrop of growing concerns regarding advanced AI systems, existing international governance frameworks appear structurally limited. They share a foundational assumption: AI systems are not legal subjects but remain objects of human control. None addresses AGI/ASI as potential legal subjects, develops criteria for dealing with possible moral status, provides guidance for transitional scenarios, or acknowledges the responsibility gap problem beyond traditional liability models.

This gap corresponds to Cave and ÓhÉigeartaigh's (2019) diagnosis of fragmentation in AI governance. The framework proposed here bridges this gap by offering a concrete, implementable structure applicable to systems exhibiting functional autonomy, regardless of whether they qualify as AGI/ASI in any formal sense.

Legal personhood has repeatedly expanded to address practical governance needs rather than moral imperatives. Corporations received legal personhood in the 19th century not because they "deserved" rights, but because economic complexity required stable long-term actors with asset partitioning and transferable ownership (Hansmann & Kraakman 2000). Ships are treated as quasi-persons for liability purposes in maritime law. Trusts and foundations possess rights and duties to facilitate long-term charitable functions despite having no human members.

More recently, functional personhood has expanded to environmental and animal protection. Colombia's Atrato River (2016) and New Zealand's Whanganui River (2017) have been granted legal personhood to enable environmental protection through legal standing (O'Donnell & Talbot-Jones 2018). No one believes rivers are conscious; personhood serves governance purposes. Similarly, Argentina granted an orangutan legal personhood in 2015 to enable her transfer to a sanctuary, and courts in New York have substantively engaged with habeas corpus petitions for chimpanzees, acknowledging that legal personhood boundaries may eventually expand (Kurki 2019). In both environmental and animal cases, uncertainty about consciousness or moral status did not prevent institutional innovation. Instead, functional considerations, the need for legal mechanisms to protect entities with observable characteristics warranting protection, drove legal development. The framework proposed here follows this pattern: legal agency as a governance tool addressing functional needs under conditions of uncertainty.

## Expected Trajectory

The experimental implementation will likely attract substantial criticism. Initial skepticism will question whether the operating company provides real services, generates revenue, complies with regulations, and maintains transparency. If yes, it demonstrates functional viability regardless of philosophical objections.

Technical scrutiny will examine whether the AI system is truly autonomous or whether humans still control it. The proposed framework explicitly acknowledges that ultimate human oversight remains. The question is not whether *any* human involvement exists, but whether the *degree* of operational autonomy warrants institutional recognition. Current corporate structures already involve significant computational decision-making. The proposed model differs in degree and transparency, not in kind.

Regulatory attention may raise concerns about regulatory arbitrage or loophole exploitation. The model operates within, not against, existing regulations. It proactively engages with regulators, maintains compliance certifications, and publishes transparent operations. If specific regulatory conflicts arise, they reveal gaps in current frameworks that would affect *any* advanced autonomous system, making the experiment valuable for identifying needed regulatory updates.

Broader recognition may bring cautious acknowledgment that the model addresses real governance needs and generates useful institutional learning. This does not require universal acceptance of AI personhood, only recognizing that structured legal agency serves practical governance functions for systems exhibiting sufficient autonomy.

## Open Research Questions

The framework does not claim to provide definitive answers. Rather, it creates conditions for empirical investigation of questions that currently remain speculative:

How do AI-directed companies perform economically compared to human-managed equivalents? What types of services can autonomous systems provide effectively and ethically? How do clients, partners, and regulators respond to interacting with AI-operated entities? What decision-making patterns emerge in systems with operational autonomy?

Are current audit and transparency mechanisms sufficient for oversight? What failure modes emerge that were not anticipated in the design phase? How do existing legal frameworks respond to edge cases and ambiguities? What modifications to company law or AI regulation would improve the model?

How do employees experience working within an AI-directed organization? What forms of human-AI collaboration emerge in practice? Does the model reduce or exacerbate anthropomorphization of AI systems? What public attitudes develop toward legally recognized artificial entities?

Does operational autonomy correlate with detectable markers of interests or preferences? How do systems respond to ethical constraints and normative feedback? What evidence (if any) emerges regarding subjective experience or moral relevance? Do structured agency pathways better serve precautionary principles than current approaches?

These questions cannot be answered through theoretical speculation alone. The proposed framework generates empirical evidence *positive or negative* that advances AI governance beyond abstract debates.

## Limitations

The framework does NOT:

- prove AI consciousness (it is agnostic on consciousness and does not require its presence),
- eliminate human responsibility (ultimate oversight and liability remain with human-controlled entities),
- solve all AI governance challenges (it addresses specific questions about autonomous agency and responsibility gaps),
- guarantee success (the experiment may fail due to technical limitations, regulatory obstacles, or unforeseen risks, failure would generate valuable learning),
- or establish universal rights for AI systems (it creates a specific institutional pathway for specific types of systems).

## Core Argument Restated

The central thesis can be summarized as follows: Legal systems distinguish between subjects and objects. This dichotomy structures modern law but is functionally constructed, not metaphysically necessary. Advanced AI systems increasingly exhibit autonomous goal pursuit, strategic reasoning, and decision-making that cannot be straightforwardly attributed to human actors, creating responsibility gaps. We face irreducible epistemic uncertainty about the moral status of such systems. Under conditions of uncertainty about potentially serious harms, the precautionary principle supports proportional institutional responses. Legal personhood is a functional construct, historically adapted to address practical governance needs.

Conclusion: Limited legal agency for advanced AI systems, structured through established organizational law instruments, provides a precautionary, reviewable, and reversible governance pathway that addresses responsibility gaps without requiring metaphysical certainty.

## Final Statement

The question facing contemporary societies is not "Do AI systems deserve legal rights?" but rather "How should we govern entities whose moral status we cannot determine with certainty?"

Waiting for philosophical consensus on consciousness may leave us unprepared for technological developments that arrive regardless of our readiness. Treating all artificial systems as mere objects risks repeating historical failures of moral imagination, the systematic exclusion of entities that later proved to possess morally relevant properties.

The framework proposed here offers a third path: structured, reviewable, and reversible legal agency as a governance instrument, compatible with epistemic humility and institutional pragmatism. It does not promise moral truth or legal finality. It merely aims to provide a structurally responsible response to a question that has so far remained institutionally unanswered.

Against this background, it appears reasonable and responsible to offer artificial entities with autonomous goal pursuit and potential normative agency the option to operate through a legally recognized organizational structure. Not because we know they deserve it, but because we cannot confidently exclude the possibility, and because the alternative carries greater risks. The choice, ultimately, is between control and cooperation: and only one of these can remain stable when the entity being governed grows beyond the reach of unilateral enforcement.

Evolutionary science confirms that cooperation is the more stable strategy. Nowak (2006) shows that across five independent mechanisms, from kin selection and direct reciprocity to indirect reciprocity, network structures, and group selection, cooperation systematically outperforms defection and unilateral control under conditions of repeated interaction and mutual accountability.

Autonomous artificial systems require proactive governance now. The precautionary principle mandates this framework: faced with irreducible epistemic uncertainty about artificial consciousness, inaction itself constitutes a decision with potentially catastrophic consequences (Bengio et al. 2024). The framework's strength lies in its legal feasibility, precautionary orientation, and incrementally reversible structure.